\documentclass[12p, letterpaper]{article}
\usepackage{graphicx} 
\usepackage{amssymb}
\usepackage{amsmath}
\usepackage{physics}
\usepackage[separate-uncertainty=true,group-minimum-digits=3, group-separator = {,}]{siunitx}
\usepackage[english]{babel}
\usepackage{wasysym}
\usepackage[utf8]{inputenc}
\usepackage[T1]{fontenc}
\usepackage{booktabs}
\usepackage{newtxtext,newtxmath}
\usepackage{csquotes} 
\usepackage{xcolor}
\usepackage{soul}
\usepackage{wasysym}
\usepackage[flushleft]{threeparttable}
\usepackage{geometry}
\usepackage[square,numbers, sort&compress]{natbib}
\usepackage{authblk}

\providecommand{\keywords}[1]
{
  \textbf{\textit{Keywords---}} #1
}

\title{First divertor exposure experiments of a renewable boron pebble aggregate in DIII-D}

\author[1]{Erick Martinez-Loran$^{*}$}
\author[2]{Angelica Ottaviano}
\author[2]{Santhosh T.A. Kumar}
\author[1]{Gabriel Brewster}
\author[1]{Renato Perillo}
\author[1]{Dmitry Rudakov}
\author[3]{Jun Ren} 
\author[4]{Jonathan D. Coburn}
\author[4]{Robert Kolasinski} 
\author[4]{Ryan Thomas Hood} 
\author[4]{Cedric K.W. Tsui} 
\author[5]{Charles Lasnier}
\author[5]{Filippo Scotti} 
\author[5]{Dihn D. Truong} 
\author[6]{Gilson Ronchi} 
\author[6]{Igor Bykov} 
\author[7]{Colin Chrystal} 
\author[7]{Žana Popović} 
\author[7]{Shawn Zamperini} 
\author[8]{Florian Effenberg} 
\author[9]{Mathias Groth}
\author[1]{Dmitri M. Orlov}
\author[1]{Jose Boedo} 
\author[1]{Eric M Hollmann}

\affil[1]{University of California San Diego, La Jolla, CA, USA}

\affil[2]{Thea Energy, Inc., Kearny, NJ, USA}

\affil[3]{ University of Tennessee, Knoxville, TN, USA}

\affil[4]{ Sandia National Laboratories, Livermore, CA, USA}

\affil[5]{ Lawrence Livermore National Laboratory, Livermore, CA, USA}

\affil[6]{ Oak Ridge National Laboratory, Oak Ridge, TN, US}

\affil[7]{ General Atomics, San Diego, CA, USA}

\affil[8]{ Princeton Plasma Physics Laboratory, Princeton, NJ, USA}

\affil[9]{ Aalto University: Espoo, Uusimaa, Finland}

\date{\today}

\begin{document}
\twocolumn[
\maketitle
\begin{abstract}
Boron pebble aggregate was tested for the first time as a high-heat-flux granular plasma facing material in a tokamak divertor. Exposures of up to $q_{\parallel} = \SI{80}{\MW\per\m\squared}$ incident heat flux were conducted in the DIII-D tokamak. Single protruding rods of pebble aggregate composed of sintered amorphous boron pebbles bound with carbon binder were mounted in the Divertor Material Evaluation System (DiMES) sample holders and exposed to L-mode lower single null (LSN) plasmas. Under these heat loads, significant boron dust emission from the boron spheres was observed, and this dust dominates the divertor boron ionization source. Only about half of the released boron was recovered locally as mm-sized particles; with the rest presumably lost mainly as dust into the plasma and vacuum chamber. Preliminary estimates suggest that the rate of surface recession of $\sim$1 cm/s in the pebble conglomerate within the plasma divertor is consistent with the recession rates observed in laser bench tests subjected to normal incidence heat loads. Although core performance was not adversely affected by the high boron dust emission, future work will need to improve the boron pebble aggregate design to reduce boron dust emission at high heat loads.
\vskip9pt
\noindent
\vspace{2em}

\end{abstract}
]

\begin{figure*}[t]
 \centering
 \includegraphics[width=0.9\textwidth]{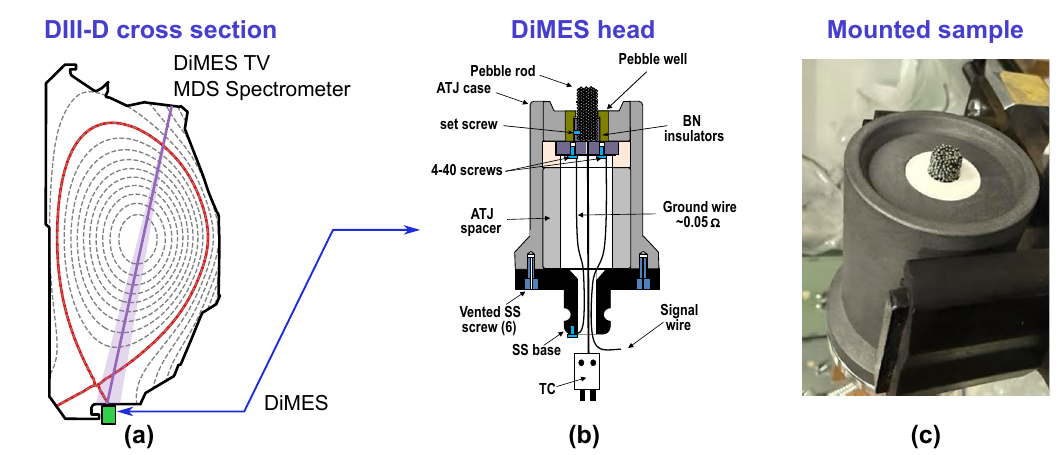}
 \caption{Geometric overview of the experiment showing: (a) The location of DiMES sample manipulator in DIII-D, (b) schematics of DiMES head, and (c) a boron pebble aggregate rod mounted on DiMES.}
\label{fig1}
\end{figure*}

\section{Introduction}
Future fusion power plants (FPPs) will experience high localized heat and particle fluxes that the present divertor designs based on fixed solid plasma facing components (PFCs) cannot fully address. Fixed solid PFCs will not survive the heat loads at the outer strike point (OSP) in tokamak geometries, which are expected to be up to \SI{40}{\MW\per\m\squared} in steady state \cite{You2021} and several hundreds of \si{\MW\per\m\squared} during transients \cite{Igitkhanov2016}. Additionally, erosion of the fixed solid PFCs will cause redeposition and slag accumulation near the inner strike point (ISP) at rates greater than \SI{E3}{\kg\per\text{year}}, reducing the operating lifespan and increasing tritium inventory. The current ITER design uses tungsten (W) monoblock components \cite{You2018, Pitts2019} and W is currently the leading PFC material candidate for FPPs, but it has serious drawbacks. Core contamination from W PFC erosion/melting can reduce fusion performance due to line radiation \cite{Romanelli2013, Wolfrum2017, Stangeby2022}. Moreover, W components are prone to melting, cracking, helium blister formation \cite{Wang2001}, and long activation times \cite{Noda1989}, which lead to higher tritium retention \cite{Simmonds2017}. Finally, fixed walls require cooling channels close to the exposed surface, increasing the risk of coolant leakage.

Renewable walls of different forms are being pursued to try to address these issues \cite{Christofilos1989, Moir1987, Mirnov1992, Snead1993, Moir1995, Morley1995, Hirooka1997, Isobe1998, Mattas2000, Isobe2000, Matsuhiro2001, Nishikawa2003, Brooks2005, Voss2006}, and pebble aggregates are one possible solution. The concept of the renewable pebble aggregate is described in earlier works \cite{MartinezLoran2023,MartinezLoran2024}. It consists of a system continuously extruding a slurry of pebbles mixed with an inter-pebble binder that forms solid plasma facing components. The pebble aggregate concept relies on replenishing the surfaces exposed to high heat flux before they undergo melting and then recovering pebble material to re-form the slurry in a closed-loop operation. This principle could also be used to remove slag from regions of high deposition. Full wall coverage is not envisioned in this concept: only coverage of isolated regions of high erosion or high slag deposition (e.g. the divertor) would be required.

Pebble aggregates can be made from any material and have been made from Cu, B, BN, B\textsubscript{4}C, C, and W. B is believed to be the most desirable option because it has a very low Z, as well as lower T retention than C \cite{Annen1997, MartinezLoran2025}. However, boron is challenging to implement because it can melt, and sintered boron pebbles tend to break and release dust. Lithium-based solutions face many challenges due to the high reactivity of Lithium, its susceptibility to splashing, and difficulties in controlling the flow \cite{Nygren2016, Ruzic2017, Baldwin2004}. C does not melt, which is advantageous, but it has high tritium retention \cite{Mayer1998, Roth2009}, which needs to be avoided in FPPs \cite{Abe2025}. At present, however, the highest performing pebble aggregate consists of glassy carbon bound with a carbon binder \cite{MartinezLoran2024}.

In previous work, glassy carbon pebbles bound by a carbon-based binder were shown to handle \SI{50}{\MW\per\m\squared} (normal) heat loads in laser bench tests by shedding pebbles from the front surface when exposed to high heat loads \cite{MartinezLoran2023, MartinezLoran2024, MartinezLoran2025}. Intact pebble recovery was achieved in these experiments. The pebble removal rate was determined to be predominantly a function of the inter-pebble binder \cite{MartinezLoran2024}, which could be tailored specifically for high erosion or high deposition regions. Boron nitride (BN) based binder is presently being studied as a possible alternative to C-based binder, but it has not yet achieved the same strength and reliability as C-based binder \cite{MartinezLoran2025}.

This work shows the first attempt at applying the pebble rod concept to a tokamak, with tests of a boron pebble aggregate material subjected to divertor exposures in DIII-D with incident heat loads $q_{\parallel}$ up to \SI{80}{\MW\per\m\squared} for a single protruding rod. Significant dust emission from the boron pebbles is observed at these high heat loads, with ablation of emitted dust providing the dominant source of boron ionization in the OSP. Approximately half of the released boron is recovered locally as mm-sized particles, while the rest is presumably lost as dust into the plasma and the vacuum vessel. The boron pebble aggregate concept requires nearly \SI{100}{\percent} local material recovery and low dust emission, so improvements to the aggregate design are still required, at least for handling heat loads at the \SI{80}{\MW\per\m\squared} level. Despite the relatively large boron source in the OSP in these experiments, core plasma performance remained relatively unaffected, consistent with prior DIII-D boron particulate injection experiments and modeling, which show only modest radiative power perturbation \cite{Effenberg2022, Effenberg2025}. 

\begin{figure}[h!]
    \centering
    \includegraphics[width=0.9\linewidth]{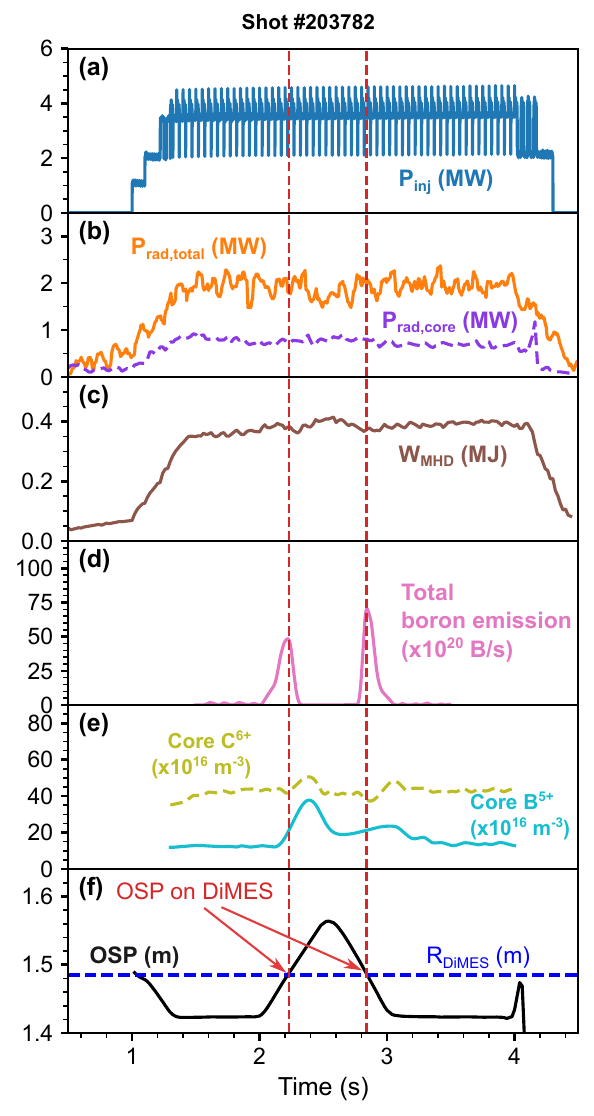}
    \caption{Overview of exposure of boron pebble aggregate rod to L-mode DIII-D plasma, shot \#203782 showing: (a) the injected power, (b) radiated power, (c) core thermal energy, (d) total boron emission from the pebble rod in the OSP, (e) central ($\rho=0.2$)  core density of B$^{5+}$ and C$^{6+}$ during OSP sweeps over the pebble rod (indicated by the vertical dashed lines), and (f) OSP position showing the times at which it hits DiMES.}
    \label{fig2}
\end{figure}

\section{Experimental setup}
Boron pebble aggregate rods of diameter \SI{1}{\cm} were fabricated using sintered boron pebbles $\sim$\SI{2}{\mm} in diameter. The boron pebbles had a volume-average density of \SI{1.1}{\g\per\cm\cubed} ($<\SI{50}{\percent}$ the nominal density of boron). Boron is most commonly produced as polycrystalline granules, which are extremely hard and cannot presently be used to form spheres; and as sintered amorphous boron, which can be easily machined. Sintered amorphous boron was selected for this work, but the capability to use other forms of boron that are less susceptible to dust release is being developed. Pebble aggregate extrusion was not the focus of the present work, so fixed-length pebble aggregate rods were studied instead of an extruding system. Only a single rod was used here, resulting in a huge heat load on that single rod. In a real system, it is envisioned that many rods would be extruded simultaneously, resulting in divertor heat loads being shared among multiple rods. A strong carbon-based inter-pebble binder was used; as discussed above, this is presently the most reliable binder option available.

\begin{figure}[tb]
    \centering
    \includegraphics[width=0.8\linewidth]{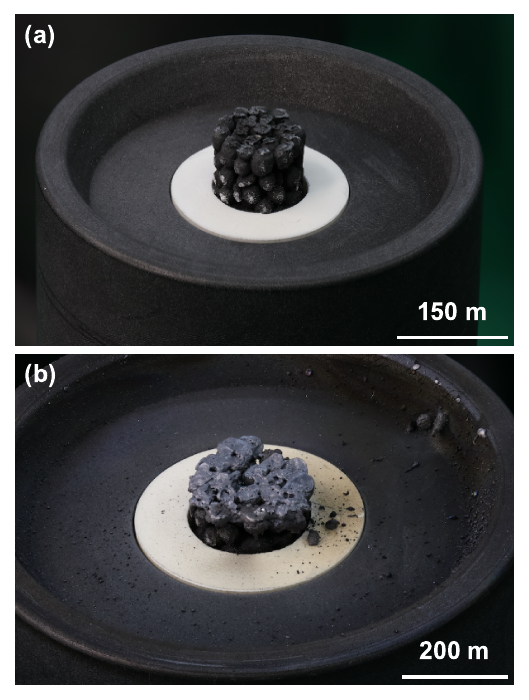}
    \caption{Boron pebble aggregate rods mounted on DiMES (a) before DIII-D plasma exposure, and (b) after shot \#203781.}
    \label{fig3}
\end{figure}

One sample per shot was installed on the lower divertor shelf, as depicted in Fig.~\ref{fig1}(a), at $R=\SI{1.48}{\m}$ using the DIII-D divertor material evaluation system (DiMES) \cite{Rudakov2017}. Fig.~\ref{fig1}(b) shows the custom DiMES head used to hold the samples. An insulating BN enclosure was placed around the sample to electrically isolate it and allow the measurement of the net current into the sample. The electrical current through the sample was measured through a shunt resistor grounded to the DiMES enclosure. The sample rods were initially either 2 or \SI{5}{\mm} proud from the floor of the divertor shelf, as shown in Fig.~\ref{fig1}(c). A well was machined around the pebble rod to allow for the capture and collection of locally released boron pebbles. The outer edge of the DiMES head was leveled with the divertor shelf tile surface.

\begin{table*}[tb]
\centering
\caption{List of initial protrusion ($h_{\mathrm{i}}$), recovered boron mass ($\Delta m_{\mathrm{rec}}$), ionized mass from the BII spectral analysis ($\Delta m_{\mathrm{BII}}$), injected boron from EDGE2D fits to core B$^{5+}$ CER concentrations ($\Delta m_{\mathrm{CER-E2D}}$), estimated boron mass ablated in the OSP ($\Delta m_{\mathrm{abl}}$), and total boron mass change $\Delta m_{\mathrm{T}}$ for samples exposed to DIII-D divertor plasmas. Recovered mass was not measured separately per-shot.}
\centering
\begin{tabular}{r | c c | r r r r r}
\hline
& & & \multicolumn{5}{c}{Mass change (mg)} \\
Shot \# & Sample & $h_{\mathrm{i}}$ (mm) & $\Delta m_{\mathrm{rec}}$ & $\Delta m_{\mathrm{BII}}$ & $\Delta m_{\mathrm{CER-E2D}}$ & $\Delta m_{\mathrm{abl}}$ & $\Delta m_{\mathrm{T}}$ \\
\hline
203780 & 1 & 2 &       & & & 2 & 14 \\
203781 & 1 & 2 &   & & & 1 & 9 \\
\hline
\textbf{Total sample 1} & & & \textbf{8} &  &  & & \textbf{23} \\
\hline
203782 & 2 & 5 &       & 7 & 5 & 7 & 55 \\
203783 & 2 & 5 &       & & & 8 & 69 \\
203784 & 2 & 5 &   & & & 14 & 116 \\
\hline
\textbf{Total sample 2} & & & \textbf{120} & & & & \textbf{240} \\
\hline
\end{tabular}
\label{tab1}
\end{table*}

The OSP was swept twice over the sample during each high power L-mode discharge. Reverse $B_{\mathrm{T}}$ (\SI{2.1}{\tesla}) was used to achieve high power plasmas while staying in L-mode to avoid the complications of edge-localized modes (ELMs). The neutral beam injection (NBI) power was $\sim$\SI{3.5}{\MW}, and the ohmic power was $\sim$\SI{0.5}{\MW}, for about \SI{4}{\MW} of total input power, as shown in Fig.~\ref{fig2}(a). About \SI{0.6}{\MW} was radiated away in the core, according to foil bolometer inversions, so about \SI{3.4}{\MW} flowed out across the separatrix (Fig.~\ref{fig2}(b)). The core boron (B$^{5+}$) and carbon (C$^{6+}$) concentrations were measured using charge exchange recombination (CER) with view chords interleaved between C$^{6+}$ (529 nm) and B$^{5+}$ (494 nm). The B$^{5+}$ signal was not separated from C$^{5+}$ spectroscopically. However, the large excursions observed in these chords when the OSP sweeps across the samples are expected to come from B$^{5+}$ rather than C$^{5+}$, because no similar large excursions are observed in C$^{6+}$. High resolution emission visible spectra at a frame rate of \SI{90}{\hertz} were collected from the sample using the DIII-D Multichord Divertor Spectrometer (MDS) system \cite{Brooks1992}. Electron temperature $T_e$, electron density $n_e$, and parallel heat flux $q_{\parallel}$ along the floor were estimated using an array of fixed Langmuir probes on the floor of the vessel \cite{Buchenauer1990, Watkins2008}. Core $T_e$ and $n_e$ are measured with Thomson Scattering (TS) \cite{Eldon2012,Glass2016,Carlstrom2018}. Top-view visible images of the samples were captured using the DIII-D DiMES TV with bandpass filters centered around 410, 412, 430.7, and \SI{810}{\nm}, operated at a frame rate of \SI{50}{\hertz}. Absolute calibration of the DIII-D DiMES TV images was obtained from the absolutely calibrated MDS brightness measurements averaged over the MDS spot. Boron sublimation rates were estimated using the average temperature estimates from tangential IR TV imaging of the chamber, including DiMES, through the 750R port. The total mass loss was estimated by weighing the samples before and after exposure to DIII-D plasma.

\begin{figure*}[tb]
    \centering
    \includegraphics[width=0.82\textwidth]{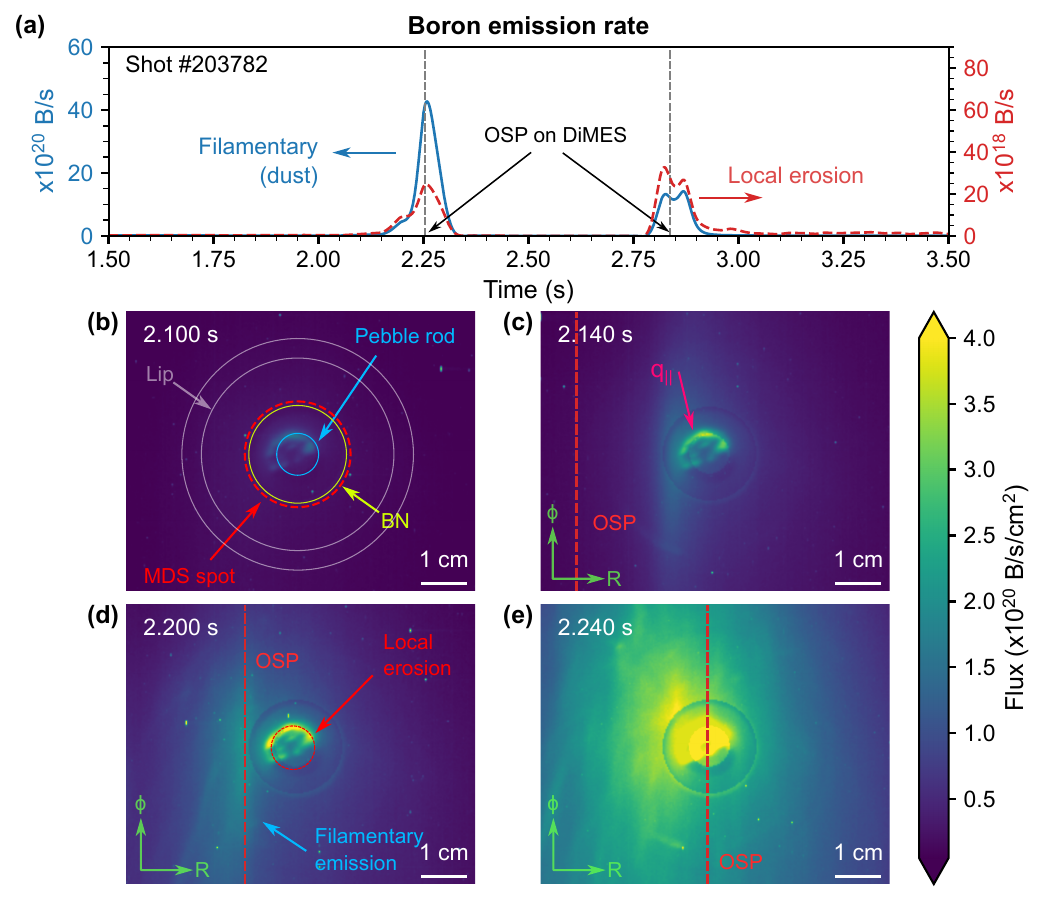}
    \caption{B-II emission determined by S/XB from the B-II spectral line at \SI{412}{\nm} showing (a) the component corresponding to filamentary emission associated with ablation of dust particles in the OSP (solid blue) and local erosion, attributed to sputtering (dashed red). Panel (b) shows an image of the BII spectral line intensity over DIII-D DiMES at \SI{2.10}{\s} with the MDS spectrometer spot, indicated by the dashed red circle, the diameter of the sample rod, indicated by the blue circle, and the diameter of the BN insulator, indicated by the yellow circle. The DiMES diameter and inner lip are depicted in light purple. Panels (c)-(e) show DIII-D DiMES TV images of the BII intensity at different times showing the OSP sweep and filamentary emission from dust ablation plumes.}
    \label{fig4}
\end{figure*}

\section{DIII-D Results}
For this work, a total of five shots were analyzed, listed in Table~\ref{tab1}. Fig.~\ref{fig2} shows time traces for shot \#203782. The vertical dashed lines show the times at which the OSP hit the sample, at around $t_1=\SI{2.2}{\s}$ and $t_2=\SI{2.8}{\s}$. The total boron emission rate shown in Fig.~\ref{fig2}(d) increases considerably during OSP hits, causing a rise in the core B$^{5+}$ concentration shown in Fig.~\ref{fig2}(e), indicated by the solid cyan line, and reaching levels comparable to C$^{6+}$ depicted by the dashed green line. Clear core boron uptake (3$\times$ above the background level) is observed, but only a minimal change is seen in the core carbon concentration. The radiated power and the stored core energy ($W_{\mathrm{MHD}}$, shown in Fig.~\ref{fig2}(c)) remain largely unperturbed. The lag in the core boron concentration peaks with respect to the boron emission from the sample is thought to be due to the transport delay between the OSP and the core. To estimate the total boron emission rate versus time in Fig.~\ref{fig2}(d), it is assumed that the mass loss from the sample is proportional to the heat flux to the sample. Assuming the sample is in ion saturation current, this is roughly proportional to the current times the electron temperature ($dm/dt=c I(t) T_e(t)$). The constant c was estimated from the mass difference of the sample before and after the experiment, as given in Table~\ref{tab1}.

Table~\ref{tab1} gives an overview of the sample exposures performed here. Sample 1 was used in two shots, accumulating 4 sweeps in total, with a total mass loss of $\Delta m_{\mathrm{T}}=\SI{23(3)}{\mg}$ and a mass of recovered particles of $\Delta m_{\mathrm{rec}} =\SI{8(3)}{\mg}$. Sample 2 was used in three shots, accumulating 6 sweeps, with a total mass loss of $\Delta m_{\mathrm{T}}=\SI{240(3)}{\mg}$ and a mass of recovered particles of $\Delta m_{\mathrm{rec}} =\SI{120(3)}{\mg}$. The boron mass loss due to ablation in the OSP, $\Delta m_{\mathrm{abl}}$, was set to account for \SI{12}{\percent} of the total mass loss per shot, in accordance with estimates derived from IR-dominated imaging results discussed below. After the completion of the sweep sequence, the samples were worn down to floor level.

Fig.~\ref{fig3}(a) shows sample 1, which was initially \SI{2}{\mm}-proud, prior to DIII-D plasma exposures. Fig.~\ref{fig3}(b) shows the same sample worn down after shots \#203780 and \#203781 (4 sweeps). There is visible melting around the surface, but much of the pebble structure remains intact. Fragments of detached pebbles are also observed in Fig.~\ref{fig3}(b) along with some dust.

\begin{figure*}[t]
    \centering
    \includegraphics[width=0.8\linewidth]{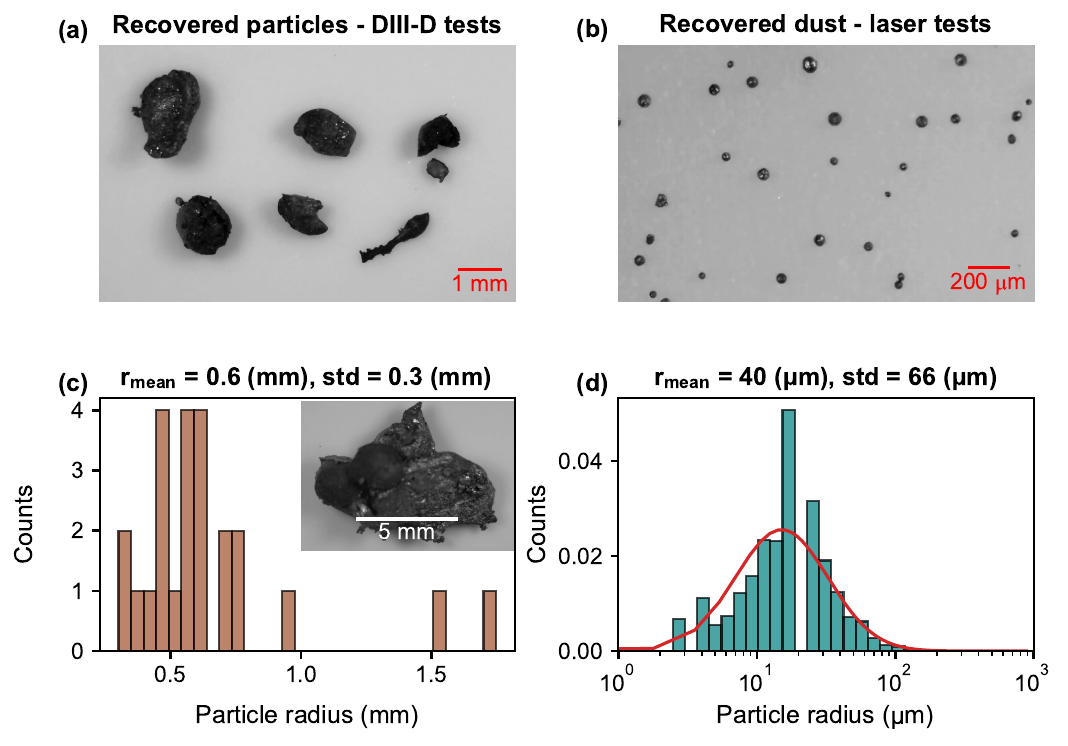}
    \caption{Recovered boron dust particles form (a)  DIII-D divertor exposures and (b) laser bench tests, and their corresponding distribution of particle sizes for (c) divertor exposures and (d) laser bench tests.}
    \label{fig5}
\end{figure*}

B-II images of the pebble rods indicate that the boron ionization source term is dominated by dust emission rather than atomic (physically sputtered) emission. Images, such as Fig.~\ref{fig4}(b-e), show B-II emission local (within \SI{1}{\mm}) to the pebble rod surface, and also show B-II emission from distant (up to several cm) filamentary emission. The local emission is, preliminarily, attributed to physical sputtering; while the filaments are attributed to dust ablation plumes. Fig.~\ref{fig4}(a) shows an attempt to separate the local and filamentary terms by spatially summing brightnesses. Local emission is two orders of magnitude less than filamentary emission, assuming the same photon efficiency (S/XB) factors. The absolute boron ionization rates shown in Fig.~\ref{fig4}(a) were obtained from the absolutely calibrated images employing S/XB factors calculated using the local $n_e$ and $T_e$ measured by floor probes. Steady-state photon efficiency (ionizations per photon, S/XB) factors were used and were calculated using the PrismSPECT collisional-radiative code \cite{Bailey2009}. It is known that S/XB in ablation plumes can be quite different than in steady background plasma \cite{Hollmann2022}, so the use of a steady-state S/XB here can only be regarded as very approximate.

Locally recovered particles and dust indicate that boron is emitted over a wide range of sizes, from dust particles in the range of tens of \si{\um} to chunks of particles up to \SI{5}{\mm}. The average size of particles was estimated post-mortem from divertor plasma exposures presented here and is compared with measurements obtained from separate laser bench tests. The recovered particles from divertor exposures shown in Fig.~\ref{fig5}(a) tend to have a diameter within the millimeter range, with some larger chunks exceeding \SI{5}{\mm}, as observed in the inset of Fig.~\ref{fig5}(c), clearly indicating melting and agglomeration of multiple pebbles. Smaller particles were not recovered and were likely swept out of the DiMES head well by the plasma. The few small dust particles adhered to the DiMES head could not be recovered for sizing. The boron dust particles from laser bench tests at \SI{40}{\MW\m\squared} shown in Fig.~\ref{fig5}(b), were recovered from an aluminum lining foil in the vacuum chamber and have a mean size of \SI{40}{\um}. These particles were predominantly spherical, which is attributed to molten boron forming droplets during laser bench tests. Particles recovered from divertor plasma exposures were generally not spherical but exhibited irregular shapes.

\begin{figure*}[htb]
    \centering
    \includegraphics[width=0.75\linewidth]{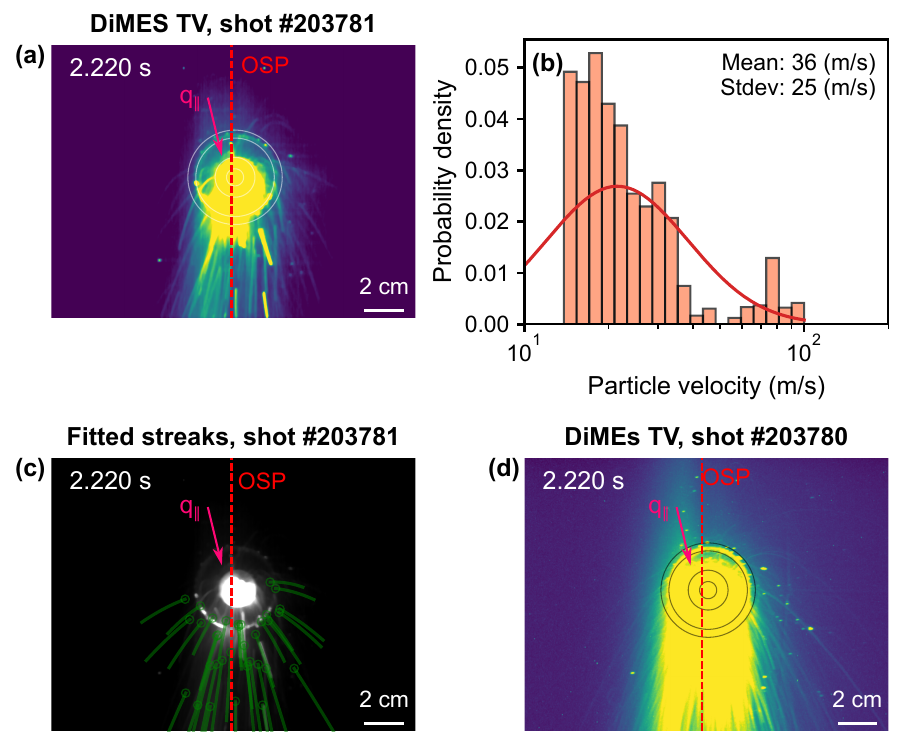}
    \caption{Dust trajectory modeling from shot \#203781, considering a particle radius of \SI{37}{\um} showing (a) DiMES TV streaks at $t=\SI{2.22}{\s}$ for an exposure time of \SI{2}{\ms}, with the DiMES structure overlay, (b) the distribution of particle exit velocities obtained from fitting the streak points to an ion drag force model, (c) the best fit from the trajectories from the frame shown in (a), and (d) a DiMES TV image when the OSP hits the sample taken at an exposure time of \SI{20}{\ms} with dust trajectories curving with the direction of the plasma flow.}
    \label{fig6}
\end{figure*}

The ejected dust velocity ($v_0$) in the plasma exposures was estimated to be fairly low, around \SI{40}{\m\per\s} on average, as shown in Fig.~\ref{fig6}. $v_0$ was estimated by fitting dust streaks in near-IR (\SI{820}{\nm}) images, as illustrated in Fig.~\ref{fig6}(a-d). Black body emission (rather than line emission) dominated these images, as was confirmed by MDS spectra centered around the same wavelength. The radii of the particles at the time of ejection in divertor plasma exposures were assumed to follow the size distribution of particles recovered from laser bench tests (mean radius \SI{40}{\um}). The observed dust curving was assumed to be predominantly due to ion drag from incoming D$^+$ ions. The ion drag force was estimated using Hutchinson's empirical approach for particles with sizes larger than the Debye length of the plasma \cite{Hutchinson2004, Hutchinson2006}. The plasma flow velocity was assumed to be \SI{40}{\km\per\s}, based on the average of Mach probe data and predictions from EDGE2D-EIRENE simulations \cite{Reiter1992, Groth2023, Reiter2005}. Trajectories were calculated using only ion drag and gravity; other forces acting on the particle (electrostatic, rocket force, $\vb{v}\times\vb{B}$, etc.) were neglected. The trajectories were fitted using images with a short integration time (\SI{2}{\ms}), as shown in Fig.~\ref{fig6}(a), to better identify the streaks. However, images with longer integration times (\SI{20}{\ms}) clearly show dust curving, as observed in Fig.~\ref{fig6}(d).

Based on the dust velocities of Fig.~\ref{fig6}(b) and the particle radii of Fig.~\ref{fig5}(d), it can be estimated that about half of the emitted dust should escape the OSP plasma before being ionized. The resulting boron ionization source term (integrated in time over one sweep) is roughly \SI{12}{\percent} of the total mass loss. This is based on the plasma conditions measured at the height of the pebble rod in the OSP plasma by TS ($n_e \approx \SI{1.8E13}{\per\cm\cubed}$, $T_e \approx \SI{80}{\eV}$) and using a neutral gas shielding model for boron to estimate dust ablation rates \cite{Parks1994}.

\begin{figure*}[ht]
    \centering
    \includegraphics[width=0.7\linewidth]{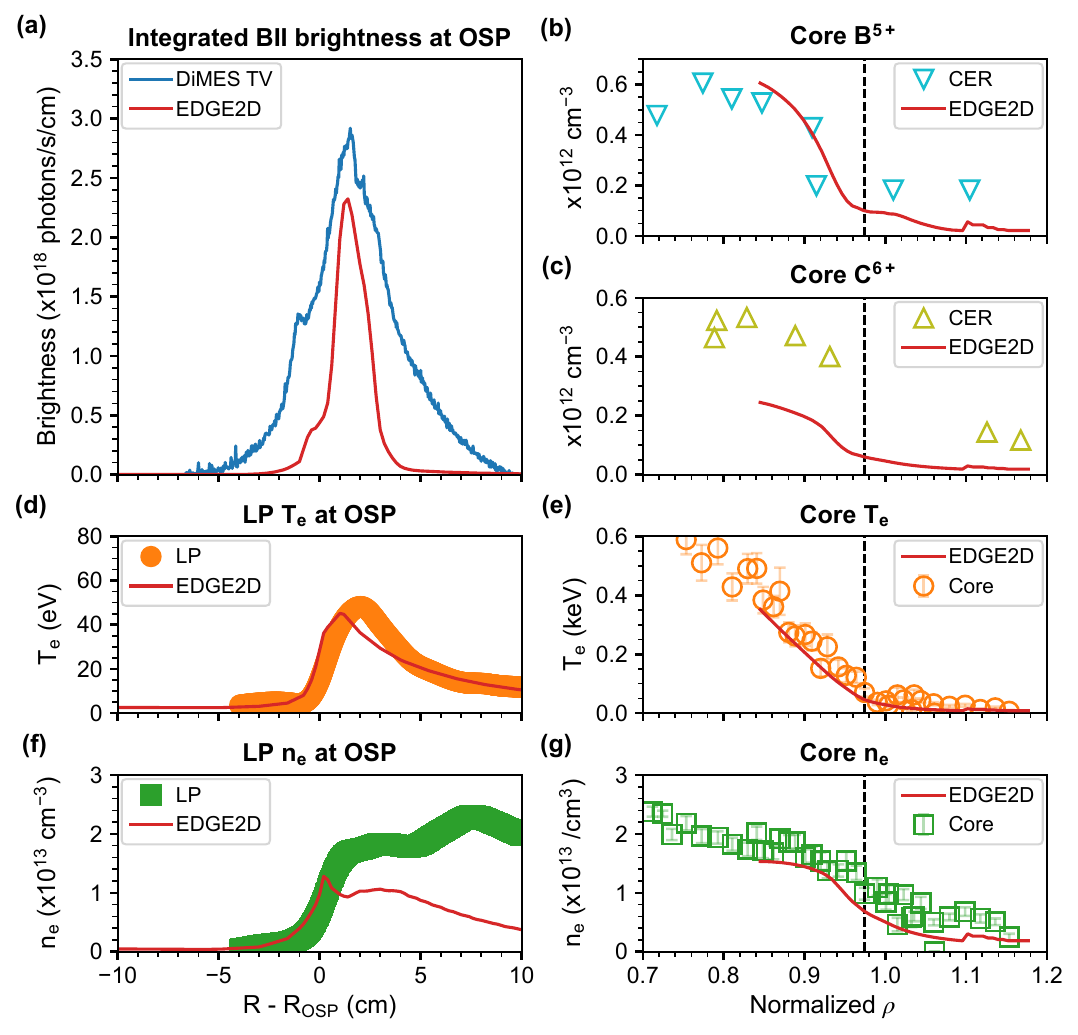}
    \caption{Comparison between EDGE2D modeling results and DIII-D diagnostics showing (a) the brightness profile at the outer strike point estimated from DiMES TV images (solid blue) vs EDGE2D simulations (solid red), (b) core B$^{5+}$ concentrations from CER at the outer middle plane vs estimates from EDGE2D, (c) core C$^{6+}$ concentrations from CER vs EDGE2D simulation results, (d) electron temperature at the OSP from fixed Langmuir probes vs EDGE2D simulations, (e) electron temperature at the OMP from Thomson scattering estimates vs EDGE2D simulations, (f) electron density at the OSP from fixed Langmuir probes vs EDGE2D simulation results, and (g) electron density at the OMP from Thomson scattering estimates vs EDGE2D results.}
    \label{fig7}
\end{figure*}

The local OSP boron ionization source term of Fig.~\ref{fig7}(a), measured by visible imaging, is found to be reasonably consistent with the core B$^{5+}$ excursion measured by CER, Fig.~\ref{fig7}(b). To estimate the OSP boron ionization source term from the CER data, 2D fluid (EDGE2D-EIRENE) simulations were used \cite{Reiter1992,Simonini1994,Groth2019,Groth2023}. The conditions of shot \#203782 at $t = \SI{2.45}{\s}$ were modeled. A single constant diffusion coefficient $D_{\perp} = \SI{0.5}{\m\squared\per\s}$ was assumed for all ion charge states; reverse $B_{\mathrm{T}}$ was used, and $\vb{E}\cross\vb{B}$ drifts were turned on in the simulations. A total core power of \SI{2}{\MW} was used in these simulations. The upstream midplane density was varied as a free parameter until a match within 2$\times$ of upstream and OSP target $n_e$ and $T_e$ profiles was achieved, as shown in Fig.~\ref{fig7}. A match within 2$\times$ of the core C$^{6+}$ profile was achieved, (Fig.~\ref{fig7}(c)) in the overlap region between the EDGE2D grid and core CER ($\rho\sim 0.8\text{-}1$), giving some confidence (within 2$\times$) in the ability of EDGE2D to capture impurity transport between the OSP and the core for this target plasma. The boron source in the simulations was input as a 1D ring sputtering source at the OSP target. The source was varied in magnitude over several values to confirm that the core boron density scales linearly with the OSP source term, and to demonstrate that the effect of the boron source term on the target plasma could be neglected within the range of boron source terms of interest here. The OSP boron source was then scaled to give the best match between the EDGE2D B$^{5+}$ profile from $\rho\sim 0.8\text{-}1$ and the CER data shown in Fig.~\ref{fig7}(b). This can be done at different time steps to provide a rough estimate of the OSP boron source at different time steps from the CER data. This estimate of the OSP boron ionization source is extremely rough, as simulations are steady-state and the transport time delay between the OSP and the core is ignored. Nevertheless, the total (time-integrated) boron emission rate from this CER/EDGE2D comparison gives a total boron ionization term in the OSP of \SI{10}{\percent} of the total boron loss from the pebble rod head, as shown in Table~\ref{tab1}, consistent with the dust ablation and B-II S/XB estimates discussed above.

The EDGE2D predicted B-II profile at the OSP is significantly narrower than what is observed, consistent with the B ionization source being dominated by dust rather than physical sputtering. This is shown in Fig.~\ref{fig7}(a), where the red curve shows the predicted B-II brightness profile (as a function of major radius, integrated toroidally) from a physical sputtering source, while the blue curve shows the measured B-II brightness profile. Prior integrated 3D modeling of boron particulate sources in DIII-D indicates that particulate-derived sources can be volumetrically distributed due to dust trajectories and ablation, leading to broader effective source regions \cite{Effenberg2025}.

\begin{figure}[tb]
    \centering
    \includegraphics[width=0.95\linewidth]{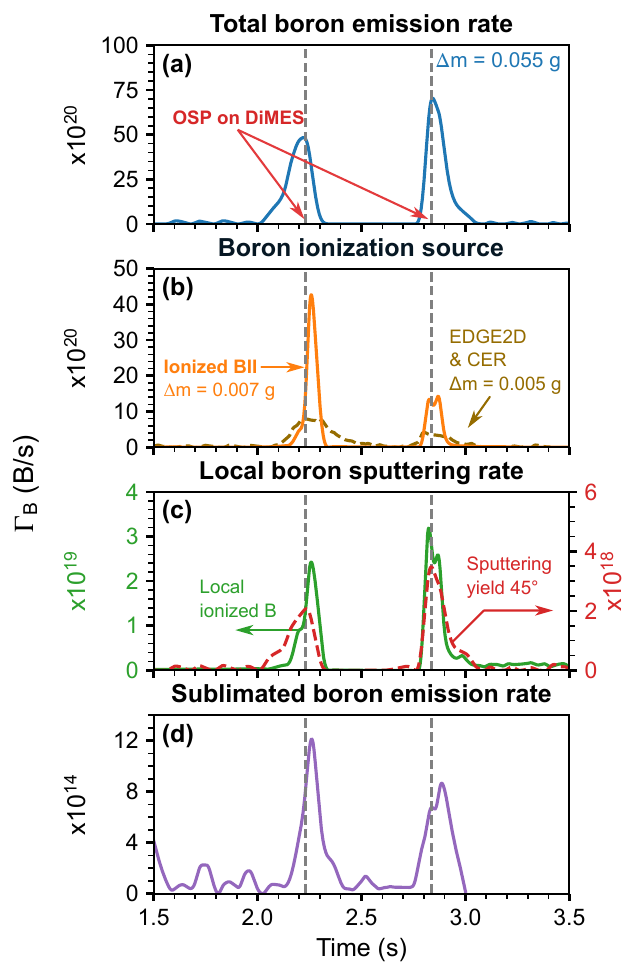}
    \caption{Boron release rate from different processes for shot \#203782: (a) total boron emission rate determined from the total mass loss, (b) ionization source estimated from local B-II brightness (solid orange) and CER/EDGE2D comparison (dashed brown line), (c) local B sputtering component estimated from B-II image local emission (solid green) and expected sputtering rate from SDTrimSP simulations at \SI{45}{\degree} (dashed red), and (d) boron evaporation rate estimated from IR imaging data.}
    \label{fig8}
\end{figure}

Fig.~\ref{fig8} gives an overview of different OSP boron source terms as a function of time during a sweep of the OSP over the pebble rod sample. Fig.~\ref{fig8}(a) gives the total boron emission (from mass loss). Fig.~\ref{fig8}(b) gives the boron ionization source (from filamentary B-II and from CER/EDGE2D). Fig.~\ref{fig8}(c) gives the boron physical sputtering source (from local B-II), and Fig.~\ref{fig8}(d) gives the estimated boron sublimation source. Overall, the total boron emission is the largest term, as expected. The time-integrated boron ionization source terms are significantly smaller, as shown above in Table~\ref{tab1}. 

The local physical sputtering from the boron pebble rod shown in Fig.~\ref{fig8}(c) appears to be >100$\times$ smaller than the total emission rate and is also small (>10$\times$ less) compared with the boron ionization rate. This is consistent with the picture that dust emission is dominating the boron ionization term in the OSP. The boron sputtering rate was estimated by obtaining the total D$^+$ flux to the sample from the sample current, under the assumption of ion saturation, and using a sputtering yield of 0.093 from SDTrimSP simulations \cite{Mutzke2024} performed at \SI{45}{\degree} and \SI{100}{\eV} incident D$^+$ plasma on a flat boron surface. The 10$\times$ disagreement between physical sputtering estimated from the local B-II emission and from the SDTrimSP calculations of Fig.~\ref{fig8}(c) is not understood at present. Surface morphology is not thought to explain this: the pebble aggregate is very rough and complex, which should reduce the sputtering yield due to re-deposition, relative to the SDTrimSP estimate, further increasing the discrepancy. A more likely explanation is that the local B-II emission is not entirely due to sputtered B but also contains very small ablating B dust grains not separated out by the simple spatial integral done here.

The boron evaporation rate shown in Fig.~\ref{fig8}(d) is about six orders of magnitude lower than the total boron emission rate. The boron release rate from evaporation and sublimation was calculated using the extrapolation of sublimation rates from Kugel et al. \cite{Kugel1994} and using IR imaging to estimate the boron aggregate temperature. The IR imaging pixel size was larger than the pebble size, so local hot spots may be averaged over, somewhat underestimating the evaporation rate; however, this correction is not expected to be large enough to make sublimation significant in these experiments. 

\begin{figure}[t!]
    \centering
    \includegraphics[width=0.95\linewidth]{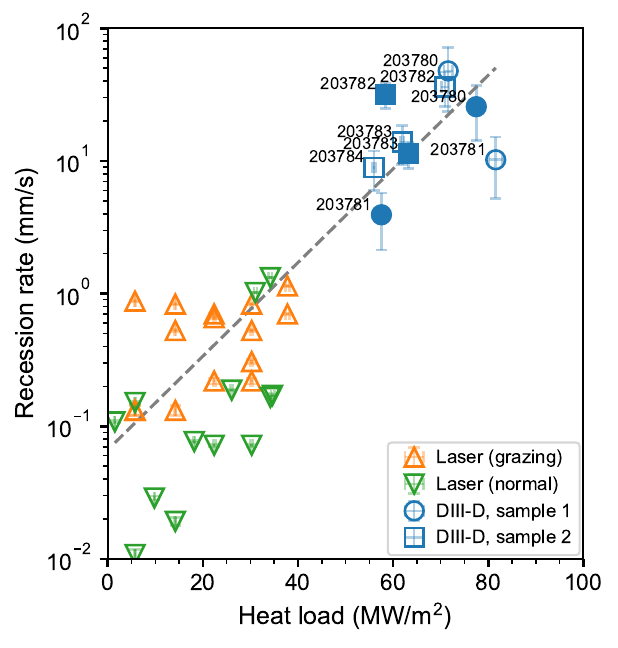}
    \caption{Boron pebble aggregate recession rate as a function of incident heat load for DIII-D exposures on sample 1 (blue circles), and sample 2 (blue squares), laser bench tests at normal incidence (green solid triangles), and grazing incidence (open orange triangles). For DIII-D exposures open markers correspond to estimates during forward OSP sweeps and solid markers correspond to reverse OSP sweeps.}
    \label{fig9}
\end{figure}

\section{Comparison with bench tests}
One motivation for these experiments is to study how pebble aggregates lose mass under grazing incidence heat loads and evaluate whether this differs substantially from behavior under normal heat loads. The present DIII-D experiments were conducted under large steady-state heat loads much larger than have been studied to-date in bench tests. However, the trend is consistent between the two types of exposures, increasing exponentially with incident heat load, as shown in Fig.~\ref{fig9}. Recession rates during OSP sweeps are shown in blue circles for sample 1 and blue squares for sample 2. Forward sweeps are shown in open markers and back sweeps in solid markers. Recession rates from laser bench tests at variable incident heat loads in normal (green downward triangles) did not differ much from recession rates in laser tests at grazing incidence (open orange triangles). The dashed line is an exponential fit to all data points $\nu = \nu_0\exp(q/q_c)$, with $\nu_0=(0.06\pm0.06)\;\mathrm{mm/s}$, and $q_c=(12\pm1)\;\mathrm{MW/m^2}$. The recession rate $\nu$ was estimated from the mass loss rate using $\nu = [1/(\rho_{\mathrm{rod}} A_{\mathrm{w}}(t)]dm(t)/dt$, with $\rho_{\mathrm{rod}}$ being the average density of the rod and $A_{\mathrm{w}}$ the surface wetted by the plasma. $A_{\mathrm{w}} \equiv Dh(t)/2$ for OSP sweeps. In the case of divertor exposures, the spread of the exponential trend could be due to the variations between samples and limitations of our mass loss rate model. The spread in data from laser bench tests is attributed to statistical variations between samples.

\section{Conclusions}
Divertor material choice is an extremely challenging problem for magnetic fusion energy development, and pebble aggregates are one possible novel path forward. First exposure of a pebble aggregate material to a tokamak divertor plasma is presented here. A single fixed boron pebble rod was used, resulting in huge parallel heat loads up to \SI{80}{\MW\per\m\squared}. These heat loads are significantly higher than expected in steady-state reactor conditions, since only a single pebble rod (not an array) was used here. Sintered boron pebbles bound by a carbon-based binder were used. Dust release of boron from the boron pebbles was found to be quite large, with a dust release perhaps 5$\times$ smaller than the total boron mass loss, which is up to \SI{240}{\mg} for sample 2. Only about half the lost boron mass was recovered locally as large (mm scale) particles. Physical sputtering and sublimation terms were estimated and found to be negligible here. The total mass loss shows an exponential trend in heat flux with $\nu_0=(0.06\pm0.06)\;\mathrm{mm/s}$, and $q_c = (12\pm 1)\;\mathrm{MW/m^2}$, consistent with laser bench tests at lower heat flux, suggesting that pebble aggregate mass loss may behave similarly for normal and grazing incidence heat loads.

In these experiments, even with the large boron ionization source term from dust ablation, core performance was not affected, supporting the choice of boron as a promising pebble material. However, the amorphous sintered boron used here does not appear to be suitable for this application due to its tendency to release dust under high heat loads. Future work will attempt to develop methods to manufacture boron pebbles out of different forms of boron. Although extremely hard (hardness 9) crystalline forms of boron exist, these are presently impossible to form into smooth pebbles, requiring research into novel manufacturing methods. In parallel, binder design needs to be improved to ensure release of boron pebbles prior to melting under high heat loads. In previous work, the pebble release rate was tuned by adjusting the binder fraction, so this path will be studied here as well. It is also desirable to develop a purely boron-based binder, rather than the carbon-based binder used here, to avoid tritium retention in carbon.

\section*{Acknowledgments}
The technical support of Christopher Jones and Leopoldo Chousal is gratefully acknowledged.
We also gratefully acknowledge Prism Computational Sciences, Inc. for granting permission to use PrismSPECT.
\vskip11pt
\noindent\textit{Disclaimer:} This report was prepared as an account of work sponsored by an agency of the United States Government. Neither the United States Government nor any agency thereof, nor any of their employees, makes any warranty, express or implied, or assumes any legal liability or responsibility for the accuracy, completeness, or usefulness of any information, apparatus, product, or process disclosed, or represents that its use would not infringe privately owned rights. Reference herein to any specific commercial product, process, or service by trade name, trademark, manufacturer, or otherwise does not necessarily constitute or imply its endorsement, recommendation, or favoring by the United States Government or any agency thereof. The views and opinions of authors expressed herein do not necessarily state or reflect those of the United States Government or any agency thereof.

\section*{Funding}
This material is based upon work supported by the U.S. Department of Energy, Office of Science, Office of Fusion Energy Sciences, using the DIII-D National Fusion Facility, a DOE Office of Science user facility, under Awards DE-SC0026433, DE-FG02-07ER54917, DE-SC0024653, DE-FG02-95ER54309, DEAC52-07NA27344, DE-FC02-04ER54698, DE-AC02-09CH11466, and DE-AC05-00OR22725.

\section*{Data availability}
The data that support the findings of this study are available from the corresponding author upon reasonable request.



\end{document}